\documentclass[12pt,a4paper,onecolumn,singlespacing,usenatbib]{mnras}
\usepackage{ae,aecompl}
\usepackage{amsmath}
\usepackage{natbib}
\title{$R_h=ct$ and the eternal coasting  cosmological model}
\author[Moncy V. John]{Moncy V. John$^{1,2}$  \thanks{moncyjohn@yahoo.co.uk}\\
$^1$ Department of Physics, St. Thomas College, Kozhencherri - 689641, Kerala, India \\
$^2$ School of Pure and Applied Physics, Mahatma Gandhi University, Kottayam, India}

\pubyear{2018}

\date{\today}

\begin{document}
\label{firstpage}
\pagerange{\pageref{firstpage}--\pageref{lastpage}}
\maketitle

\begin{abstract}
 We point out that the nonempty $R_h=ct$ cosmological model has some known antecedents in the literature. Some of those eternal coasting models are published even before the discovery of the accelerated expansion of the universe and were shown to have none of the commonly discussed cosmological problems and also  that  $H_0t_0=1$. The $R_h=ct$ model is only the special (flat) case of the eternal coasting model. An additional feature in the coasting model is that  $\Omega_m/\Omega_{dark \; energy}$ = some constant of the order of unity, so that also the cosmic coincidence problem is avoided. 

\end{abstract}


\section{Introduction}

A cosmological model  named   `$R_h=ct$ model' was proposed and studied  by F. Melia and collaborators  \citep{melia2007,melia2012rh}. Observational and theoretical aspects of this model, which claims to be a new one,  are discussed in several papers \citep{melia2012fitting,melia2013,melia2013cosmic,melia2014,melia2015recent,melia2015cosmological,
melia2016epoch,melia2016galaxy,melia2018maximum}.

Here I wish to point out that this model is identical   to a pre-existing `eternal coasting cosmological model'  \citep{mvjkbj2000}, which is a non-empty (non-Milne)  one. The latter  was  studied at length by J.V. Narlikar, K. Babu Joseph and the present author in several publications \citep{mvjvn2002,mvj2005,mvj2015}. Between these two models, there exist only some minor differences,  and even these  find their origin in  certain vague and unsupported features  assumed in the $R_h=ct$ model. The  pre-existing model in \citep{mvjkbj2000} itself had some antecedents \citep{mvjkbj1996,mvjkbj1997}, which predicted a  nonempty (non-Milne) coasting universe excepting the Planck epoch. This  was proposed even before the release of the sensational SNe Ia data in 1998. The authors of the $R_h=ct$ model have performed almost the same kind of data analysis  as in our case and reaches  similar conclusions. Though these differently named models originated on   different theoretical grounds,  they all end up at  the same cosmological solution for all practical purposes.

After a brief review of the two models in the following two sections, we present  our specific contentions on this issue in the last section.

In these discussions, we shall not be concerned with how  the authors of the papers on $R_h=ct$ cosmological model  and the eternal coasting model arrived at their models. Rather, we shall concentrate on the resulting cosmological model and its observational predictions.

\section{The $R_h=ct$ model}
 The authors of the $R_h=ct$ model' \citep{melia2007,melia2012rh} have described their model as one like a Friedman-Robertson-Walker (FRW)  model which assumes the presence of dark energy as well as matter and radiation \citep{melia2015cosmological,melia2015recent,melia2016epoch,melia2016galaxy,melia2017}. The $\Lambda$CDM model  is the mostly favoured such model.  But the principal difference between this and the $R_h=ct$ model is that the latter is  constrained by the equation of state $\rho+3p=0$ (the so-called active mass condition in general relativity), in terms of the total pressure $p$ and energy density $\rho$ of the universe. 

Thus they have

\begin{equation}
\rho = \rho_m+\rho_r+\rho_{de}, \label{eq:rho}
\end{equation}

\begin{equation}
p=p_m+p_r+p_{de}, \label{eq:p}
\end{equation}
and

\begin{equation}
\rho+3p=0. \label{eq:rhoplus3p}
\end{equation}
Here $\rho_m$, $\rho_r$ and $\rho_{de}$ are the densities of matter, radiation and dark energy, respectively, in the universe and $p_m=0$, $p_r=(1/3)\rho_r$ and $p_{de}=w_{de}\rho_{de}$ are the corresponding pressures.
Since general relativity is assumed valid, the total energy in this FRW model is conserved by virtue of a continuity equation to be obeyed by the total energy-momentum tensor. However, matter, radiation and dark energy are not assumed to be separately  conserved in the $R_h=ct$ model \citep{melia2015recent}. 

The authors make several additional assumptions, such as flat geometry \citep{melia2013cosmic}, constancy of the equation of state parameter for dark energy $w_{de}\equiv p_{de}/\rho_{de} \approx -0.5$, etc. (In fact, they admit that $w_{de}$ need not be a constant in their model, but prefer to call the latter  an assumption for simplicity \citep{melia2016epoch}.) The assumption of flat geometry is also somewhat arbitrary, for there is no genuine need for a mechanism such as inflation in this case.

While comparing the performance of their model with the $\Lambda$CDM model, the authors have stated explicitly the predictions of the $R_h=ct$ model. The luminosity distance in this model is given by \citep{melia2016galaxy}

\begin{equation}
D_L^{R_h=ct} =\frac{c}{H_0}(1+z)\ln (1+z), \label{eq:luminosityd}
\end{equation}
and the angular diameter distance is given as \citep{melia2018maximum}

\begin{equation}
d_A^{R_h=ct}(z)=\frac{c}{H_0}\frac{1}{(1+z)}\ln (1+z). \label{eq:angdimd}
\end{equation}
They have also  stated that for applications such as these, one does not need to know the detailed make-up of the cosmic fluid. In summary, for the $R_h=ct$ model it is immaterial  how the various energy components $\rho_m$, $\rho_r$, $\rho_{de}$, etc., balance each other to make  $\rho+3p=0$, at least in  cases such as the above cosmological observations. These authors have never outlined any explicit mechanism or time dependence for these individual components, with which this balancing can be achieved.

It may be noted that \cite{melia2016epoch} have stated that in the $R_h=ct$ model,

\begin{equation}
\rho^{R_h=ct}=\rho_c (1+z)^2.
\end{equation}
The reason for this  is ascribed to the feature that all of the energy components must together produce a total equation of state $p=-(1/3) \rho$ in this model. 
But conspicuously, in spite of the voluminous work done by them, the authors working in this field have not    explicitly stated that their model requires

\begin{equation}
\rho = \rho_m+\rho_r+\rho_{de} \propto \frac{1}{a^2}, \label{eq:rhoprop}
\end{equation}
which  follows directly from the conservation equation for total energy

\begin{equation}
\frac{d}{da}(\rho a^3)+3pa^2=0,
\end{equation}
in all FRW models. How they can account for this simple but mandatory variation  of the total energy density with scale factor as in Eq. (\ref{eq:rhoprop}), without causing a coincidence problem, is of utmost importance.

\section{The Eternal Coasting cosmology}
The antecedents to the $R_h=ct$ cosmology in \citep{mvjkbj1996,mvjkbj1997,mvjkbj2000} arrive at the non-empty `always coasting' model (scale factor $a\propto t$ or equivalently $\rho+3p=0$) by starting from different premises. These are also  FRW models which contain matter, radiation and dark energy, but it is more general than the $R_h=ct$ because the discussions are not restricted  to a flat geometry for the universe. (We take $k=-1$, $0$, or $+1$ respectively  for open, flat and closed geometries in the RW metric.) In \citep{mvjkbj2000}, a dimensional argument led to the postulate that in the classical epoch of the universe (i.e.,when the scale factor $a\gg l_p$, the Planck length), all energy densities must vary as $a^{-2}$. Similar evolution is predicted in \citep{mvjkbj1996,mvjkbj1997} too. Thus in this model, in which equations (\ref{eq:rho})-(\ref{eq:rhoplus3p}) are assumed valid and which obeys equations of state $p_m=0$, $p_r=(1/3)\rho_r$ and $p_{de}=w_{de} \rho_{de}$, one postulates

\begin{equation}
\rho_m \propto a^{-2}, \label{eq:rho_mprop}
\end{equation}

\begin{equation}
\rho_p \propto a^{-2},  \label{eq:rho_pprop}
\end{equation}

\begin{equation}
\rho_{de} \propto a^{-2}, \label{eq:rho_deprop}
\end{equation}
so that also Eq. (\ref{eq:rhoprop}) is valid here. These variations are possible since the individual components need not be separately conserved. (In the above papers, the model was presented as a decaying-$\Lambda$ cosmology with creation of matter/radiation.) The evolution of the scale factor is then obtained as $a=mt$, with $m$ as a constant.   This evolution gives for the flat case, 

\begin{equation}
R_h=\frac{c}{H_0}= ct,
\end{equation}
which is the same result obtained in the `$R_h=ct$' model. But we may obtain a more general relation in all cases of open, closed and flat geometries as 

\begin{equation}
H_0 t_0=1.
\end{equation}
 This  equation is  more characteristic for the model. If we assume the dark energy to be due to a cosmological constant, then $p_{de}=-\rho_{de}$. A very interesting prediction of the model is then

\begin{equation}
\frac{\rho_r}{\rho_{de}}=\frac{\Omega_r}{\Omega_{de}}=1,   \label{eq:ratio1}
\end{equation}
in the early relativistic era, and

\begin{equation}
\frac{\rho_m}{\rho_{de}}=\frac{\Omega_m}{\Omega_{de}}=2,  \label{eq:ratio2}
\end{equation}
in the late nonrelativistic era. These   predictions are made in \citep{mvjkbj1996,mvjkbj1997,mvjkbj2000}, where $w_{de}=-1$.  For a specific instance, see Eq. (7) in \citep{mvjkbj2000}. In this model, the ratio of density parameters of matter and dark energy shall remain a constant of the order of unity for all times and this is not a feature that appeared in the present universe only. (In the $\Lambda$CDM model, the near equality of these densities is only for the present epoch and it is referred to as the `coincidence problem'.) 

It is worth mentioning that one can adjust the value of $w_{de}$ (by making use of the mysterious nature of dark energy) to get  the ratio of density parameters, etc. at some other desired values.  If dark energy is with $p_{de}=-(1/2)\rho_{de}$, then  the eternal coasting model also predicts ${\rho_m}/{\rho_{de}}={1}/{2}$. However, one must note that such values for the ratio of density parameters obtained from the analysis of cosmological data are highly model-dependent \citep{mvj2004}. The values obtained in the coasting model (or $R_h=ct$ model) need not be compared with  $\Omega_m \approx 0.28$ and $\Omega_{\Lambda}\approx 0.72$, because the latter are values obtained in analyses which assume the $\Lambda$CDM model at the outset.

In summary, the near equality of matter/radiation and dark energy densities is a prediction in the eternal coasting model, and this helps to avoid the cosmic `coincidence problem'. In addition, all other commonly discussed cosmological problems such as flatness, horizon, age, etc. are absent in this model, as discussed at length in the above papers.

\section{Identical models}

Delimiting the eternal coasting model to  flat geometry alone does not help to make a new model. There is no justification for delimiting the model to the flat geometry either. Even when one considers all the three geometries (open, closed and flat), there is no place for any flatness problem in the eternal coasting model. Hence there is no genuine need for a mechanism such as inflation in this case. 

In this section, we present two  arguments to show that the $R_h=ct$ model is not a new one and is only a special case of the nonempty eternal coasting model in \citep{mvjkbj1996,mvjkbj1997,mvjkbj2000}.

(1) At the observational front,   the luminosity distance and the angular diameter distance are two  important observable quantities. In the $a=mt$ coasting model, the luminosity distance is given by the expression \citep{mvjvn2002,mvj2005}

\begin{equation}
D_L^{coasting}=\frac{c}{H_0}m(1+z)\sin n \left[ \frac{1}{m}\ln (1+z)\right], \label{eq:luminosityd_coast}
\end{equation}
and the angular diameter distance by

\begin{equation}
d_A^{coasting}=\frac{c}{H_0}\frac{m}{(1+z)} \sin n \left[\frac{1}{m} \ln (1+z) \right]. \label{eq:angdimd_coast}
\end{equation}
Here all the three geometries are considered, with $\sin n (x) = \sinh (x)$ for open geometry,  $\sin n (x) = \sin (x)$ for closed geometry and  $\sin n (x) = x$ for flat geometry. We have  used these predictions of the eternal coasting ($a=m t$) model  in references \citep{mvjvn2002,mvj2005}, while performing Bayesian model comparison    with the $\Lambda$CDM model.  For instance, see the last equation in Sec. III and also Eq. (17) in \citep{mvjvn2002}.

 One can now compare the luminosity distance and the angular diameter distance for the $k=0$ flat case in Eqs. (\ref{eq:luminosityd_coast}) and  (\ref{eq:angdimd_coast}), respectively,  of the eternal coasting model with that in the $R_h=ct$  model, given by the above Eqs. (\ref{eq:luminosityd}) and  (\ref{eq:angdimd}). These are seen identical to each other.

This establishes that the $R_h=ct$ model is only a special case of the eternal coasting model, as far as the above observations are concerned.

\medskip

(2) The only provision  for the `$R_h=ct$ model'  to claim any  difference  with the `eternal coasting model' is with regard to variation of the densities of individual cosmic fluids. The authors of the former have stated that various constituents of the total energy  can  adjust their relative densities via particle-particle interactions \citep{melia2015recent} to get  evolution equations that satisfy $p=-(1/3)\; \rho$. However,  no specific time-evolution for the densities of these constituents [similar to that in equations (\ref{eq:rho_mprop}), (\ref{eq:rho_pprop}), (\ref{eq:rho_deprop})] is  suggested in the $R_h=ct$ papers. The authors excuse themselves by stating that when the evolution of individual components is needed, several conservation laws and reasonable assumptions delimit their behaviour \citep{melia2016epoch}. Hence there is no model here to compare with the eternal coasting model.  

\medskip

The above two specific  arguments are sufficient to show that the `$R_h=ct$ model' is a special case of the eternal coasting cosmological model,  published in a new name.

Lastly, we wish to add one more point: Any attempt to describe a time evolution different from that in equations (\ref{eq:rho_mprop}), (\ref{eq:rho_pprop}), (\ref{eq:rho_deprop}) will only aggravate the cosmic coincidence problem.     The problem with  any such modification  can be  seen  more clearly  by noting that since $\rho=\rho(a)$, the densities $\rho_m$, $\rho_r$, $\rho_{de}$, etc. are all functions of $a$. By definition, $\rho_m$ and $\rho_r$ are positive at all cosmic epochs and obey $p_m=0$ and $p_r=(1/3)\rho_r$.  For   coasting models, $\rho+3p=0$ or $\rho \propto a^{-2}$. Considering the fact that the individual components need not be conserved and also that these energy densities are  functions of $a$, let us assume for them  some Laurent series expansion in $a$, about $a=0$. It may then be noted that

\begin{equation}
\rho_m =\sum_{n=-\infty}^{+\infty}c_n^m\; a^n \geq 0,
\end{equation}

 \begin{equation}
\rho_r =\sum_{n=-\infty}^{+\infty}c_n^r\; a^n \geq 0,
\end{equation}

\begin{equation}
\rho_{de} =\sum_{n=-\infty}^{+\infty}c_n^{de}\; a^n,
\end{equation}
and 

\begin{equation}
\rho =\rho_m+\rho_r+\rho_{de}=\sum_{n=-\infty}^{+\infty}c_n\; a^n =c_{-2}\; a^{-2}.
\end{equation}
Equating the coefficients of equal powers of $a$ in this equation, one can see that 

\begin{equation}
c_{-2}^m+c_{-2}^r+c_{-2}^{de}=c_{-2} \geq 0, \qquad \hbox {for} \qquad n = -2,
\end{equation}
and 

\begin{equation}
c_n^m+c_n^r+c_n^{de}= 0, \qquad \hbox {for} \qquad n \neq -2. \label{eq:notequal}
\end{equation}
Here we have used the fact that since  the conserved total energy density $\rho$ needs to be positive at present, it is positive at all other epochs. Similarly at any epoch, $\rho_m$ and $\rho_r$ must be positive by definition. Then, unless we choose $c_n^m= 0$, $c_n^r= 0$, $c_n^{de}= 0$ for all $n \neq -2$,  the above condition (\ref{eq:notequal}) requires  $\rho_{de}$  to become negative at least for some other early or late epochs. A dark energy that flips sign  will only aggravate the coincidence problem. Note that all these problems are avoided in the eternal coasting model, which has the variations given by   equations (\ref{eq:rho_mprop}), (\ref{eq:rho_pprop}), (\ref{eq:rho_deprop}).

\label{lastpage}
\end{document}